\documentclass[10pt,a4paper,oneside]{article}

\usepackage{amssymb}
\usepackage{amsmath}
\usepackage{amsthm}

\newtheorem{theorem}{Theorem}[section]
\newtheorem{lemma}[theorem]{Lemma}
\theoremstyle{definition}
\newtheorem{definition}[theorem]{Definition}

\newcommand{\arch}{\mathrm{arch}}
\newcommand{\di}{\underline{d_{1}}}
\newcommand{\dii}{\underline{d_{2}}}
\newcommand{\diii}{\underline{d_{3}}}
\newcommand{\Dom}{\mathrm{Dom}}
\newcommand{\Diag}{\mathop{\textrm{Diag}}\nolimits}
\newcommand{\dint}{\mathop{\textrm{d}}\nolimits}
\newcommand{\dinti}{\mathop{\textrm{d}^{2}}\nolimits}
\newcommand{\diff}[2]{\dfrac{\dint #1}{\dint #2}}
\newcommand{\diffi}[2]{\dfrac{\dinti #1}{\dint #2}}
\newcommand{\pd}[2]{\dfrac{\partial #1}{\partial #2}}

\begin{document}

\title{Description of the free motion with momentums in G\"{o}del's universe
  \thanks{keywords: G\"{o}del universe, geodesic line; MSC: 53C22, 83C10. }}

\author{
  Bal\'{a}zs B\'{a}r\'{a}ny\footnote{
    Institute of Mathematics, BUTE, Budapest, Egry J. u. 1., 1111, Hungary.
    Email: balubsheep@gmail.com} 
  \ and 
  Attila Andai\footnote{
    RIKEN, BSI, Amari Research Unit, 2--1, Hirosawa, Wako, Saitama 351-0198, Japan.
    Email: andaia@math.bme.hu}}

\date{July 25, 2008}

\maketitle

\begin{abstract}
We study the geodesic motion in G\"{o}del's universe, using conserved quantities.
We give a necessary and sufficient condition for curves to be geodesic curves in terms of
  conserved quantities, which can be computed from the initial values of the curve.
We check our result with numerical simulations too.
\end{abstract}

\maketitle

\section{Introduction}

G\"{o}del's solution \cite{godel3} for the Einstein equation gives a cosmological model of a 
  rotating universe.
This solution has many interesting properties, for example it contains closed timelike curves,
  geodesically complete, and has neither a singularity nor a horizon.
The free motion of particles in this cosmological model was analyzed first by 
  Kundt \cite{kundt} and Chandrasekhar and Wright \cite{geod}.
A physical model for the geodesic motion was given by Novello et. al. \cite{novello} using
  the effective potential.
In this paper we describe the free motion with conserved quantities. 

In the flat spacetime $(\mathbb{R}^{4},g=\Diag(1,-1,-1,-1))$ the equation for a geodesic
  curve $\gamma:\mathbb{R}\to\mathbb{R}^{4}$ is given by
\begin{equation}
\ddot{\gamma}=0.
\end{equation}
The solution is obvious $\gamma(u)=\underline{v}t+\underline{x}$.
The vectors $\underline{v},\underline{x}$ are given as initial values.
In this case we can consider $\underline{v}$ as a conserved quantity, since a curve
  $\gamma$ describes a free motion if and only if $\dot{\gamma}(u)=\underline{v}$
  for every $u\in\Dom(\gamma)$.
In this setting the physical meaning of $\underline{v}$ is well-known.
In the next section we give a similar description for geodesic motion in the 
  G\"{o}del spacetime and we check our result using numerical simulations.
During the computation we use the Einstein summation and the usual differential geometrical
  formalism.

\section{Momentums in the G\"{o}del's universe}

\subsection{Geodesic lines}

G\"{o}del presented the line-element of his universe in the following form 
  \cite{godel3}, \cite[p. 195]{godel1}, \cite[p. 275]{godel2}.

\begin{definition}
We call the Riemannian manifold $(\mathcal{M},g)$ to \emph{G\"odel spacetime}, where
  the manifold is $\mathcal{M}$ expressed in cylindrical coordinates
  $\mathbb{R}\times\mathbb{R}^{+}\times\left\lbrack 0,2\pi\right\lbrack\times\mathbb{R}$ and the
  Riemannian metric at a point $(t,r,\varphi,z)\in\mathcal{M}$ is
\begin{equation}
g(t,r,\varphi,z)=4a^{2}\begin{pmatrix} 
  1 & 0 & \sqrt{2}\sinh^{2}r & 0 \\
  0 & -1 & 0 & 0 \\
  \sqrt{2}\sinh^{2}(r) & 0 & \sinh^{4}(r)-\sinh^{2}(r)& 0 \\
  0 & 0 & 0 & -1
\end{pmatrix},
\end{equation}
  where $a$ is positive parameter.
\end{definition}

The parameter $a$ can be interpreted as $a=\frac{1}{\sqrt{2}}\omega$,
  where $\omega$ is a measure of the constant rotation \cite[p. 191]{godel1}.

To get the differential equations of the geodesic curve we compute the second order 
  Christoffel symbols, using the Equation
\begin{equation}
\Gamma_{ij}^{m}=\frac{1}{2}g^{km}
  \left(\partial_{i}g_{jk}+\partial_{j}g_{ik}-\partial_{k}g_{ji} \right),
\end{equation}
  where $g^{km}$ denotes the $(k,m)$ element of the inverse matrix of $g$.
The nonzero symbols are the following (and their counterparts $\Gamma_{ij}^{k}=\Gamma_{ji}^{k}$).
\begin{align}
& \Gamma_{12}^{1}=\frac{2\sinh(r)}{\cosh(r)}
 & & \Gamma_{23}^{1}=\frac{\sqrt{2}\sinh^{3}(r)}{\cosh(r)}\\
& \Gamma_{13}^{2}=\frac{\sqrt{2}\sinh(r)\cosh(r)}{r^{2}}
 & & \Gamma_{23}^{1}=\frac{\sinh(r)\cosh(r)(2\cosh^{2}(r)-3)}{r^{2}}\\
& \Gamma_{12}^{3}=-\frac{\sqrt{2}}{\sinh(r)\cosh(r)}
 & & \Gamma_{23}^{3}=\frac{1}{\sinh(r)\cosh(r)}
\end{align}

A curve $\gamma:\mathbb{R}\to M$ is a geodesic line if
\begin{equation}
\ddot{\gamma}^{k}+\Gamma_{ij}^{k}\dot{\gamma}^{i}\dot{\gamma}^{k}=0
\end{equation}
  holds for $k=1,2,3,4$.
If we write the curve as $\gamma(u)=\left( t(u), r(u), \varphi(u), z(u)\right)$
  then $\gamma$ is a geodesic line if and only if the following equations hold.
\begin{align}
&\label{eq:geodline:tt}\ddot{t}+\frac{4\sinh(r)}{\cosh(r)}\dot{t}\dot{r}
  +\frac{2\sqrt{2}\sinh^{3}(r)}{\cosh(r)}\dot{\varphi}\dot{r}=0\\[1em]
&\label{eq:geodline:rr}\ddot{r}+2\sqrt{2}\sinh(r)\cosh(r)\dot{t}\dot{\varphi}+
  \sinh(r)\cosh(r)(2\cosh^{2}(r)-3)(\dot{\varphi})^{2}=0\\[1em]
&\label{eq:geodline:pp}\ddot{\varphi}\sinh(r)\cosh(r)-2\sqrt{2}\dot{t}\dot{r}
  +2\dot{\varphi}\dot{r}=0\\[1em]
&\label{eq:geodline:zz}\ddot{z}=0,
\end{align}
  where the dot $\dot{}$ denotes the derivation with respect to $u$.
It is known from the theory of differential equations that, there exists a unique solution
  for every initial value \cite{semi}.

\subsection{Description of the free motion with conserved quantities}
 
In this subsection we derive a $4\times 4$ matrix $M$, which mixes the components of 
  $\dot{\gamma}$, where $\gamma$ is a curve in the G\"{o}del spacetime, 
  such that $\gamma$ is a geodesic curve, if and only if $M\dot{\gamma}$ is constant.
Since during the geodesic motion $M\dot{\gamma}$ is constant, like the impulse momentum in
  the classical case, we call this matrix $M$ to momentum matrix.
First we define this matrix, and then we prove its mentioned properties.

\begin{definition}
For parameters $\rho_{1},\rho_{2}\in\mathbb{R}$, $r\in\mathbb{R}^{+}$ and
  $\varphi\in\left[0,2\pi\right[$ let us define the \emph{momentum matrix}
  $M(\rho_{1},\rho_{2},r,\varphi)$
\begin{equation}
\begin{pmatrix}
\rho_{2}+\dfrac{\rho_{1}\cosh^{2}(r)}{2} & 0 
  & \dfrac{(\rho_{1}\cosh^{2}(r)+4\rho_{2})\sinh^{2}(r)}{2\sqrt{2}} & 0\\[1em]
-\sqrt{2}\cos(\varphi)\sinh(2r) & \sin(\varphi) 
  & -\dfrac{\cos(\varphi)(\cosh(2r)-2)\sinh(2r)}{2} & 0 \\[1em]
\sqrt{2}\sin(\varphi)\sinh(2r) & \cos(\varphi) 
  & \dfrac{\sin(\varphi)(\cosh(2r)-2)\sinh(2r)}{2} & 0 \\[1em]
0 & 0 & 0 & 1
\end{pmatrix}.
\end{equation}
\end{definition}

\begin{theorem}
For every geodesic line $\gamma(u)=(t(u), r(u), \varphi(u),z(u))$ and for every real parameters 
  $\rho_{1}, \rho_{2}$ and  there exist real parameters $c_{1}, c_{2},c_{3}$ 
  and $f(\rho_{1},\rho_{2})$ depending just on the initial conditions $\gamma(u_{0})$ and
  $\dot{\gamma}(u_{0})$, such that
\begin{equation}
\label{eq:cons:theorem}
M(\rho_{1},\rho_{2},r,\varphi)\cdot
   \begin{pmatrix} \dot{t}        \\ \dot{r} \\ \dot{\varphi} \\ \dot{z} \end{pmatrix}
  =\begin{pmatrix} f(\rho_{1},\rho_{2}) \\ c_{1}   \\ c_{2}         \\ c_{3}   \end{pmatrix}
\end{equation}
  holds at every point of $\Dom\gamma$.
If a curve $\gamma:\mathbb{R}\to M$ satisfies Equation (\ref{eq:cons:theorem}) at
  every point of $\Dom\gamma$, then $\gamma$ is a geodesic curve.
\end{theorem}

\begin{proof}
Since the Theorem is obviously valid for the last component $z$, we skip it during the proof.
Consider the left hand side of the differential Equations 
  (\ref{eq:geodline:tt}, \ref{eq:geodline:rr}, \ref{eq:geodline:pp}) as a vector
  $\underline{X}=\left(\mbox{(\ref{eq:geodline:tt})},\mbox{(\ref{eq:geodline:rr})},
  \mbox{(\ref{eq:geodline:pp})}\right)$.
Assume that $D(t,r,\varphi)$ is a $3\times 3$ is an invertible matrix, which entries are
  smooth functions of $t,r$ and $\varphi$. 
Assume that the equation
\begin{equation}
\label{eq:conserv:pre}
\diff{}{u}\Bigl(D(t(u),r(u),\varphi(u))\cdot\dot{\gamma}(u)\Bigr)=
  D(t(u),r(u),\varphi(u))\cdot\underline{X(u)}
\end{equation}
holds.

If $D$ consists of the column-vectors $\di,\dii$ and $\diii$, then the 
  Equation (\ref{eq:conserv:pre}) can be written as
\begin{align}
&\di\ddot{t}+\dii\ddot{r}+\diii\ddot{\varphi}+\pd{\di}{t}(\dot{t})^{2}
  +\pd{\dii}{r}(\dot{r})^{2}+\pd{\diii}{\varphi}(\dot{\varphi})^{2}+\\ 
&\quad+\left(\pd{\di}{r}+\pd{\dii}{t}\right)\dot{t}\dot{r}
  +\left(\pd{\di}{\varphi}+\pd{\diii}{t}\right)\dot{t}\dot{\varphi}
  +\left(\pd{\dii}{\varphi}+\pd{\diii}{r}\right)\dot{\varphi}\dot{r}=\nonumber\\ 
&=\di\ddot{t}+\dii\ddot{r}+\diii\ddot{\varphi}
  +\sinh(r)\cosh(r)(2\cosh^{2}(r)-3)\dii(\dot{\varphi})^{2}+\nonumber\\ 
&\quad+\left(4\frac{\sinh(r)}{\cosh(r)}\di-\frac{2\sqrt{2}}{\sinh(r)\cosh(r)}\diii\right)
  \dot{t}\dot{r}
  +2\sqrt{2}\sinh(r)\cosh(r)\dii\dot{t}{u}\dot{\varphi}+\nonumber\\
&\quad+\left(2\sqrt{2}\frac{\sinh^{3}(r)}{\cosh(r)}\di+\frac{2}{\sinh(r)\cosh(r)}\diii\right)
  \dot{\varphi}\dot{r}.\nonumber
\end{align}
After elementary calculations we have the set of differential equations for the vectors  
  $(\underline{d_{i}})_{i=1,2,3}$.
\begin{eqnarray}
&&\label{eq:cons:1t}\pd{\di}{t}=0\\[1em]
&&\label{eq:cons:1r}\pd{\di}{r}=-\pd{\dii}{t}-\frac{2\sqrt{2}}{\sinh(r)\cosh(r)}\diii
  +4\di\tanh(r)\\[1em]
&&\label{eq:cons:1p}\pd{\di}{\varphi}=-\pd{\diii}{t}+\frac{2\sinh(2r)}{\sqrt{2}}\dii\\[1em]
&&\label{eq:cons:2r}\pd{\dii}{r}=0\\[1em]
&&\label{eq:cons:2p}\pd{\dii}{\varphi}=-\pd{\diii}{r}+\frac{2}{\sinh(r)\cosh(r))}\diii
  +2\sqrt{2}\cosh(r)\sinh(r)\di-2\sqrt{2}\di\tanh(r)\\[1em]
&&\label{eq:cons:3p}\pd{\diii}{\varphi}=\frac{1}{2}\dii(\cosh(2r)-2)\sinh(2r)
\end{eqnarray}
Let us note, that the same set of differential equations are valid for all components
  of the vectors $\di, \dii, \diii$.
Therefore it is enough to consider just the first component of the vectors, which will be
  denoted with $d_{1},d_{2}$ and $d_{3}$.
(That is for index $i\in\{1,2,3\}$, $d_{i}$ denotes the first component of the vector 
  $\underline{d_{i}}$.)

Equations (\ref{eq:cons:1t},\ref{eq:cons:2r}) imply that $d_{1}=d_{1}(r,\varphi)$ and
  $d_{2}=d_{2}(t,\varphi)$. 
If we compute the derivative of the Equation (\ref{eq:cons:1p}) with respect to $\varphi$
  and $t$, using Equation (\ref{eq:cons:3p}) we have the following equation for $d_{2}$.
\begin{equation}
\left(\frac{\cosh(2r)-2}{2}\right)\dfrac{\partial^{2}d_{2}}{\partial t^{2}}=
  \sqrt{2}\ \dfrac{\partial^{2}d_{2}}{\partial t \partial \varphi}
\end{equation}
Computing the derivative of the previous Equation with respect to $r$, and taking into account
  the Equation (\ref{eq:cons:2r}) we have the following form for $d_{2}$.
\begin{equation}
d_{2}(t,\varphi)=b_{1}(\varphi)t+b_{2}(\varphi)
\end{equation}
Using this form for $d_{2}$ Equation (\ref{eq:cons:3p}) gives
\begin{equation}
d_{3}(t,r,\varphi)=\frac{1}{2}\left(t\int b_{1}(\varphi)\dint\varphi
  +\int b_{2}(\varphi)\dint\varphi\right)\cdot (\cosh(2r)-2)\sinh(2r)+e(r,t),
\end{equation}
  where $e(r,t)$ is a smooth function.
Substituting this form of $d_{3}$ to the Equation (\ref{eq:cons:1r}) and computing the derivative
  with respect $t$, we have
\begin{equation}
0=-2\sqrt{2}(\cosh(2r)-2)\int b_{1}(\varphi)\dint\varphi-\frac{2\sqrt{2}}{\cosh(r)\sinh(r)}
  \pd{e(r,t)}{t}.
\end{equation}
This means, that there exists a real number $\lambda\in\mathbb{R}$, such that
\begin{equation}
\int b_{1}(\varphi)\dint\varphi=\lambda,\quad \textnormal{and}\quad
 \frac{1}{(\cosh(2r)-2)\cosh(r)\sinh(r)}\ \pd{e(r,t)}{t}=-\lambda
\end{equation}
  holds.
This means that $b_{1}(\varphi)=0$.
From this $d_{2}=b_{2}(\varphi)$ and $e(r,t)=g(r)$ follows.
So far, we have the following forms for the functions $(d_{i})_{i=1,2,3}$.
\begin{eqnarray}
&&\label{eq:cons:d1form} d_{1}=\sqrt{2}\sinh(2r)\int b_{2}(\varphi)\dint\varphi+f(r)\\[1em]
&&\label{eq:cons:d2form}d_{2}=b_{2}(\varphi)\\[1em]
&&\label{eq:cons:d3form}d_{3}=\frac{1}{2}\left(\int b_{2}(\varphi)\dint\varphi\right)
  (\cosh(2r)-2)\sinh(2r)+g(r)
\end{eqnarray}
Where Equation (\ref{eq:cons:d1form}) comes from Equation (\ref{eq:cons:1p}).
Let us substitute the Equations (\ref{eq:cons:d1form},\ref{eq:cons:d2form},\ref{eq:cons:d3form}) 
  into the Equation (\ref{eq:cons:2p}).
\begin{equation}
\begin{split}
&\diff{b_{2}}{\varphi}=-\frac{1}{2}\left(\int b_{2}(\varphi)\dint\varphi\right)
  \Bigl(2(\cosh(2r)-2)\cosh(2r)+2\sinh^{2}(2r)\Bigr)-\diff{g}{r}(r)\\[1em]
&\qquad+\frac{2}{\cosh(r)\sinh(r)}
  \left(\frac{1}{2}\left(\int b_{2}(\varphi)\dint\varphi\right)(\cosh(2r)-2)\sinh(2r)
  +g(r)\right)\\[0.5em]
&\qquad+2\sqrt{2}\left(\sqrt{2}\sinh(2r)\int b_{2}(\varphi)\dint\varphi+f(r)\right)
  \cosh(r)\sinh(r)\\[0.5em]
&\qquad-2\sqrt(2)\left(\sqrt{2}\sinh(2r)\int b_{2}(\varphi)\dint\varphi
  +f(r)\right)\frac{\sinh(r)}{\cosh(r)}.
\end{split}
\end{equation}
After simplifications we get the following equation.
\begin{equation}
\begin{split}
&\diff{b_{2}(\varphi)}{\varphi}+\int b_{2}(\varphi)\dint\varphi \\
&= -\diff{g(r)}{r}+\frac{2g(r)}{\cosh(r)\sinh(r)}
  +2\sqrt{2}f(r)\cosh(r)\sinh(r)-2\sqrt{2}f(r)\tanh(r)
\end{split}
\end{equation}
This implies that
\begin{equation}
\diff{b_{2}(\varphi)}{\varphi}+\int b_{2}(\varphi)\dint\varphi=\alpha
\end{equation}
  holds, for a real constant $\alpha$. 
This gives the explicit form of $b_{2}$
\begin{equation}
b_{2}(\varphi)=k_{1}\cos(\varphi)+k_{2}\sin(\varphi),
\end{equation}
  where $k_{1}, k_{2}\in\mathbb{R}$.
After this form of $b_{2}$ we rewrite the Equations  
  (\ref{eq:cons:d1form},\ref{eq:cons:d2form},\ref{eq:cons:d3form}).
\begin{eqnarray}
&&\label{eq:cons:d1for}d_{1}=\sqrt{2}\Bigl(k_{1}\sin(\varphi)-k_{2}\cos(\varphi)\Bigr)\sinh(2r)
  +f^{*}(r)\\[0.5em]
&&\label{eq:cons:d2for}d_{2}=k_{1}\cos(\varphi)+k_{2}\sin(\varphi)\\[0.5em]
&&\label{eq:cons:d3for}d_{3}=\frac{1}{2}\Bigl(k_{1}\sin(\varphi)-k_{2}\cos(\varphi)\Bigr)
  \Bigl(\cosh(2r)-2\Bigr)\sinh(2r)+g^{*}(r)
\end{eqnarray}
If we substitute Equations (\ref{eq:cons:d1for},\ref{eq:cons:d2for},\ref{eq:cons:d3for})
  into Equation (\ref{eq:cons:1r}), after some simplification we have
\begin{equation}
\label{eq:cons:f*}
\diff{f^{*}}{r}(r)=\frac{-4\sqrt{2}g^{*}(r)}{\sinh(2r)}+4f^{*}(r)\tanh(r).
\end{equation}
And if we substitute Equations (\ref{eq:cons:d1for},\ref{eq:cons:d2for},\ref{eq:cons:d3for})
  into Equation (\ref{eq:cons:2p}), then after simplifications we get
\begin{equation}
\label{eq:cons:g*}
\diff{g^{*}}{r}(r)=\frac{4(g^{*}(r)+\sqrt{2}f^{*}(r)\sinh^{4}(r)}{\sinh(2r)}.
\end{equation}
If we express the function $g^{*}(r)$ from Equation (\ref{eq:cons:f*}) and
  substitute it to the Equation (\ref{eq:cons:g*}) then we get a solvable differential equation.
\begin{equation}
2\cosh(2r)\diff{f^{*}}{r}(r)=\sinh(2r)\diffi{f^{*}}{r^{2}}(r)
\end{equation}
The general solution is
\begin{equation}
f^{*}(r)=\rho_{2}+\frac{1}{2}\rho_{1}\cosh(r)^{2},
\end{equation}
  where $\rho_{1},\rho_{2}\in\mathbb{R}$. 
After this, we can compute the function $g^{*}$.
\begin{equation}
g^{*}(r)=\frac{(\rho_{1}+8\rho_{2}+\rho_{1}\cosh(2r))\sinh^{2}(r)}{4\sqrt{2}}.
\end{equation}
We determined the first components of the vectors $\di,\dii$ and $\diii$.
This means, that the column-vectors of the matrix $D$ are of the following form.
\begin{eqnarray}
&&\di=\sqrt{2}(\underline{k_{1}}\sin(\varphi)-\underline{k_{2}}\cos(\varphi))\sinh(2r)
  +\underline{1}(\rho_{2}+\frac{1}{2}\rho_{1}\cosh^{2}(r))\\[0.5em]
&&\dii=\underline{k_{1}}\cos(\varphi)+\underline{k_{2}}\sin(\varphi)\\[0.5em]
&&\diii=\frac{1}{2}(\underline{k_{1}}\sin(\varphi)-\underline{k_{2}}\cos(\varphi))
  (\cosh(2r)-2)\sinh(2r)+\\[0.5em]
&&\qquad +\underline{1}\frac{(g_{1}+8g_{2}+g_{1}\cosh(2r))\sinh^{2}(r)}{4\sqrt{2}},\nonumber
\end{eqnarray}
  where $\underline{1}=(\frac{1}{\sqrt{3}},\frac{1}{\sqrt{3}},\frac{1}{\sqrt{3}})$ and the vectors 
  $\underline{k_{1}}$, $\underline{k_{2}}$ and $\underline{1}$ are linearly independent, since we
  assumed, that the matrix $D$ is invertible.
Moreover we can assume that the vectors $\underline{k_{1}}$, $\underline{k_{2}}$ 
  and $\underline{1}$ form an orthonormal basis in $\mathbb{R}^{3}$.
According to this, if we multiply from left the matrix $D$ with the matrix
  $L$ which has row-vectors $(\underline{1},\underline{k_{2}},\underline{k_{1}})$
  we get the matrix $M(\rho_{1},\rho_{2},r,\varphi)$, which is independent of $t$.
This proves the Theorem.
\end{proof}

\subsection{Numerical verification of the Theorem}
Now we give a numerical verification of the previous Theorem.
First we note that enough to consider two linearly independent constants vectors
  $(\rho_{1},\rho_{2})\in\mathbb{R}^{2}$ and $(\rho'_{1},\rho'_{2})\in\mathbb{R}^{2}$ 
  in the matrix $M$, since easy to check that the function $f(\rho_{1},\rho_{2})$ is linear
\begin{equation}
f(\rho_{1},\rho_{2})+f(\rho'_{1},\rho'_{2})=f(\rho_{1}+\rho'_{1},\rho_{2}+\rho'_{2}).
\end{equation}
We choose the matrices $M(0,1,t,r,\varphi)$ and $M\left(1,-\dfrac{1}{4},t,r,\varphi\right)$.
For these $M$ matrices the Equation (\ref{eq:cons:theorem}) can be written in the
  following form. 
(We skip the $z$ component, since this part of the Theorem is obvious.)
\begin{eqnarray}
&&\label{eq:cons:num1}\dot{t}(u)+\sqrt{2}\sinh^{2}(r)\dot{\varphi}(u)=f(0,1)\\[0.5em]
&&\label{eq:cons:num2}\frac{\cosh(2r)}{4}\dot{t}(u)+\frac{\sinh^{4}(r)}{2\sqrt{2}}\dot{\varphi}(u)=
  f\left(1,-\frac{1}{4}\right)\\[0.5em]
&&-\sqrt{2}\cos(\varphi)\sinh(2r)\dot{t}(u)+\sin(\varphi)\dot{r}(u)\nonumber\\[0.5em]
&&\label{eq:cons:num3}\qquad-\frac{\cos(\varphi)(\cosh(2r))-2)\sinh(2r)}{2}\dot{\varphi}(u)
  =c_{1}\\[0.5em]
&&\sqrt{2}\sin(\varphi)\sinh(2r)\dot{t}(u)+\cos(\varphi)\dot{r}(u)\nonumber\\[0.5em]
&&\label{eq:cons:num4}\qquad+\frac{\sin(\varphi)(\cosh(2r)-2)\sinh(2r)}{2}\dot{\varphi}(u)=c_{2}
\end{eqnarray}
Since the set of differential equations for the geodesic line is invariant with respect
  to the translation for variables $t$, $z$ and $\varphi$, we can assume that
  $t(0)=0$, $\varphi(0)=0$ and $z(0)=0$.
We choose the other initial values for the geodesic lines randomly.
For example we can choose the initial values $r(0)=1.33$, $\dot{r}(0)=0.78$,
  $\dot{t}(0)=0.81$ and $\dot{\phi}(0)=0.56$.
From these values we have the exact numbers $f(0,1)$, $ f\left(1,-\frac{1}{4}\right)$,
  $c_{1}$ and $c_{2}$ from Equations (\ref{eq:cons:num1}-\ref{eq:cons:num4}).
We use the MAPLE software, \textit{dsolve} package, \textit{dverk78} method to solve
  numerically the geodesic equations with these initial values on the interval 
  $\left[0,2\right]$.
We set the absolute and relative error for the solution to be $10^{-8}$.
We compute the quantities of the left hand side in the Equations 
  (\ref{eq:cons:num1}-\ref{eq:cons:num4}) at the points $u=k\cdot 0.005$ ($1\leq k\leq 400$),
  and we compare the computed values to the exact ones.
In this case the biggest error for Equation (\ref{eq:cons:num1}) is $E_{1}=0.5\times 10^{-8}$,
  for Equation (\ref{eq:cons:num2}) is $E_{2}=0.5\times 10^{-8}$,
  for Equation (\ref{eq:cons:num3}) is $E_{3}=0.4\times 10^{-7}$ and 
  for Equation (\ref{eq:cons:num4}) is $E_{4}=0.54\times 10^{-8}$.
The results of $20$ simulations are shown in Table~\ref{sim:table}.

\begin{table}[ht]
\caption{}\label{sim:table}
\noindent\[
\begin{array}{|c|c|c|c|c|c|c|c|}
\hline
r(0) & \dot{r}(0) & \dot{t}(0) & \dot{\phi}(0)  &
  E_{1}\times 10^{8} & E_{2}\times 10^{8} & E_{3}\times 10^{8} & E_{4}\times 10^{8} \\ \hline
 1.33 &  0.78 &  0.81  &  0.56 &  0.5  &  0.5  &  4    &  0.54 \\ \hline
 1.14 & -0.21 &  0.26  & -1.37 &  0.7  &  0.6  &  2.5  &  0.42 \\ \hline
 0.5  & -0.97 &  0.67  &  0.14 &  0.25 &  0.13 &  0.73 &  3    \\ \hline
 0.68 & -1.05 &  0.97  & -0.34 &  0.2  &  0.13 &  6.4  &  4    \\ \hline
 0.1  &  1.43 &  1.29  & -0.37 &  0.4  &  0.19 &  0.38 &  0.8  \\ \hline
 0.84 & -0.56 &  1.36  &  1.11 &  0.2  &  0.2  &  0.8  &  0.28 \\ \hline
 0.62 &  0.96 &  0.3   & -0.69 &  0.07 &  0.04 &  26.13&  6    \\ \hline
 1.26 &  0.13 &  0.75  &  0.37 &  0.3  &  0.3  &  2    &  0.2  \\ \hline
 0.29 &  0.32 &  1.26  &  0.67 &  0.1  &  0.1  &  0.15 &  0.06 \\ \hline
 0.11 & -0.52 &  1.23  & -0.03 &  0.1  &  0.1  &  0.1  &  0.11 \\ \hline
 1.28 & -0.31 &  0.67  & -0.58 &  0.4  &  0.26 &  1.7  &  0.6  \\ \hline
 0.14 & -0.82 &  1.4   & -0.51 &  0.1  &  0.2  &  0.17 &  0.19 \\ \hline
 0.35 &  0.92 &  0.02  &  0.43 &  0.04 &  0.02 &  5.32 &  1    \\ \hline
 1.24 &  0.79 &  0.69  &  1.25 &  0.8  &  0.8  &  5    &  0.8  \\ \hline
 0.15 & -0.17 &  1.49  &  0.14 &  0.1  &  0.1  &  0.15 &  0.3  \\ \hline
 0.3  &  0.45 &  0.35  &  0.97 &  0.09 &  0.06 &  0.11 &  0.28 \\ \hline
 0.41 &  0.69 &  1.49  & -0.08 &  0.1  &  0.1  &  0.4  &  0.19 \\ \hline
 0.7  & -0.2  &  1.41  &  1.48 &  0.1  &  0.2  &  0.06 &  0.04 \\ \hline
 0.2  &  0.17 &  1.14  & -1.06 &  0.1  &  0.1  &  0.12 &  0.04 \\ \hline
 0.36 & -0.84 &  1.31  & -1.08 &  0.2  &  0.1  &  0.3  &  0.18 \\ \hline
\end{array}
\]
\end{table}

It is clear from the table, that the errors are acceptable with respect to the
  computational precision.
This gives a numerical verification of the Theorem.

\subsection{Some simple consequences}

If we choose the parameters in the matrix $M$ to be $(\rho_{1},\rho_{2})=(2,0)$ and  
  $(\rho_{1},\rho_{2})=(0,1)$ then the first components of the 
  Equation (\ref{eq:cons:theorem}) are
\begin{eqnarray}
&&\dot{t}+\frac{\sinh^{2}(r)}{\sqrt{2}}\dot{\varphi}=\frac{f(2,0)}{\cosh^{2}(r)}\\[0.5em]
&&\dot{t}+\sqrt{2}\sinh^{2}(r)\dot{\varphi}=f(0,1).
\end{eqnarray}
Solving these Equations for $\dot{t}$ and $\dot{\varphi}$ we have
\begin{eqnarray}
&&\label{eq:cons:soltp}
  \dot{t}=\dfrac{2f(2,0)}{\cosh^{2}(r)}-f(0,1)\\[0.5em]
&&\label{eq:cons:solpp}
  \dot{\varphi}=\frac{\sqrt{2}}{\sinh^{2}(r)}\left(f(0,1)-\frac{f(2,0)}{\cosh^{2}(r)} \right).
\end{eqnarray}
So the functions $\dot{t}$ and $\dot{\varphi}$ depend just on the initial conditions
  and on $r$.
Moreover we can define a critical radius $R_{t}=\arch\sqrt{\frac{2f(2,0)}{f(0,1)}}$ 
  (if exists) from the initial conditions, where $\dot{t}=0$.
If $r>R_{t}$ then the particle moving backward in time and if $r<R_{t}$ then moving forward.

This leads us to the following Lemma.

\begin{lemma}
If $f(2,0)\geq 0$ and $f(0,1)\leq 0$ then $\dot{t}(u)\geq 0$ for every 
  $u\in\Dom\gamma$.
If $f(2,0)\leq 0$ and $f(0,1)\geq 0$ then $\dot{t}(u)\leq 0$ for every
  $u\in\Dom\gamma$.
\end{lemma}

Let us compute the inverse of the matrix $M$
\begin{equation}
M^{-1}=\frac{1}{\kappa}\begin{pmatrix}
8-4\cosh(2r) & -\dfrac{\sigma\cos(\varphi)\tanh(r)}{\sqrt{2}} 
  & \dfrac{\sigma\sin(\varphi)\tanh(r)}{\sqrt{2}} & 0  \\[1em]
0 & \kappa\sin(\varphi) & \kappa\cos(\varphi) & 0\\[1em]
8\sqrt{2} & \dfrac{2\sigma\cos(\varphi)}{\sinh(2r)} 
  & \dfrac{-2\sigma\sin(\varphi)}{\sinh(2r)} & 0\\[1em]
0 & 0 & 0 & \kappa
\end{pmatrix},
\end{equation}
  where
\begin{equation}
\kappa=\rho_{1}+(\rho_{1}+4\rho_{2})\cosh(2r)\quad\mbox{and}
  \quad \sigma=\rho_{1}+8\rho_{2}+\rho_{1}\cosh(2r).
\end{equation}
From these Equations proves the following Lemma.

\begin{lemma}
We have correspondences between the functions $r$ and $\varphi$
\begin{align}
& \label{eq:cons:solrp2} \dot{r}=c_{1}\sin(\varphi)+c_{2}\cos(\varphi)\\
& \label{eq:cons:solpp2} \dot{\varphi}=\frac{2}{\sinh(2r)}
  \left(c_{1}\cos(\varphi)-c_{2}\sin(\varphi) \right)+f(2,0)\frac{2\sqrt{2}}{\cosh^{2}(r)}.
\end{align}
Which means that the function $\dot{r}$ is bounded
\begin{equation}
\left|\dot{r}\right|\leq\sqrt{c_{1}^{2}+c_{2}^{2}}
\end{equation}
  and we have a bound for $\dot{\varphi}$
\begin{equation}
\left|\dot{\varphi}\right|\leq \frac{2\sqrt{c_{1}^{2}+c_{2}^{2}}}{\sinh(2r)}
  +f(2,0)\frac{2\sqrt{2}}{\cosh^{2}(r)}.
\end{equation}
\end{lemma}

From Equations (\ref{eq:cons:solpp},\ref{eq:cons:solrp2}) we get
\begin{equation}
\diff{r}{\varphi}=\frac{c_{1}\sin(\varphi)+c_{2}\cos(\varphi)}
  {\dfrac{\sqrt{2}}{\sinh^{2}(r)}\left(f(0,1)-\dfrac{f(2,0)}{\cosh^{2}(r)} \right)}.
\end{equation}
After integration we have the following Lemma.

\begin{lemma}
There is a real parameter $m$ such that
\begin{equation}
c_{2}\sin(\varphi)-c_{1}\cos(\varphi)=
  \frac{2\sqrt{2}}{\sinh(2r)}\left(f(2,0)\cosh(2r)-f(0,1)\cosh^{2}(r) \right)+m
\end{equation}
  holds.
\end{lemma}

\bigskip

{\bf Acknowledgement.}
The second author was supported by Japan Society for the Promotion of Science, 
  contract number P 06917.


\begin{thebibliography}{99}

\bibitem[1]{geod} S. Chandrasekhar, J. P. Wright, 
\textit{The geodesics in G\"odel's universe}, 
Proc. Nat. Acad. Sci. U.S.A. {\bf 47} (1961), 341--347.

\bibitem[2]{godel3} K. G\"{o}del,
\textit{An example of a new type of cosmological solutions of Einstein's field equations of
  gravitation},
Rev. Mod. Phys. \textbf{21} (1949), 447--450.

\bibitem[3]{godel1} K. G\"{o}del,
\textit{Collected Works. Volume II. Publications 1938--1974.}
Oxford University Press, 1990.
Editors: S. Feferman, J. W. Dawson Jr., S. C. Kleene, G. H. Moore, R. M. Solovay 
  and J. van Heijenoort.

\bibitem[4]{godel2} K. G\"{o}del,
\textit{Collected Works. Volume III. Unpublished Essays and Lectures.}
Oxford University Press, 1995.
Editors: S. Feferman, J. W. Dawson Jr., W. Goldfarb and C. Parsons.

\bibitem[5]{visual} A. Hajnal, J. Madar\'{a}sz, I. N\'{e}meti,
\textit{Visualizing some ideas about G\"odel-type rotating universes}, preprint 2002.

\bibitem[6]{kundt} W. Kundt,
\textit{Tr\"{a}gheitsbahnen in einem von G\"{o}del angegebenen kosmologischen modell},
Zeitschrift f\"{u}r Physik, \textbf{145} (1956), 611--620.

\bibitem[7]{novello} M. Novello, I. Dami\~{a}o Soares  and J. Tiomno,
\textit{Geodesic motion and confinement in G\"odel's universe},
Phys. Rev. D \textbf{27} (1983), 779--788.

\bibitem[8]{semi} B. O'Neill,
\textit{Semi-Riemannian Geometry},
Academic Press, New York, 1983.

\end{thebibliography}
\end{document}